\documentclass[9pt,conference]{IEEEtran}
\usepackage[preprint]{waspaa25}

\usepackage{bm} % for bold math symbols (incl. Greek letters) with \bm{}

\title{Recomposer: Event-roll-guided generative audio editing}

\name{
\begin{tabular}{c}
Daniel P. W. Ellis\textsuperscript{*}\thanks{* Core contributors.} \quad Eduardo Fonseca\textsuperscript{*} \quad Ron J. Weiss\textsuperscript{*} \quad Kevin Wilson\textsuperscript{*} \quad Scott Wisdom\textsuperscript{*}
\\
Hakan Erdogan \quad John R. Hershey \quad Aren Jansen \quad R. Channing Moore \quad Manoj Plakal
\vspace{-1mm}
\end{tabular}}

\address{Google DeepMind\\
 \normalsize
  \texttt{\{dpwe,efonseca,ronw,kwwilson,scottwisdom\}@google.com}
}

\begin{document}

\maketitle

\begin{abstract}
Editing complex real-world sound scenes is difficult because individual sound sources overlap in time.
Generative models can fill-in missing or corrupted details based on their strong prior understanding of the data domain.
We present a system for editing individual sound events within complex scenes able to delete, insert, and enhance individual sound events
based on textual edit descriptions (e.g., ``enhance Door'') and a graphical representation of the event timing 
derived from an ``event roll'' transcription.  We present an encoder-decoder transformer working on SoundStream representations,
trained on synthetic (input, desired output) audio example pairs formed by adding isolated sound events to dense, real-world backgrounds.  Evaluation reveals the importance of each part of the edit descriptions -- action, class, timing.  Our work demonstrates ``recomposition'' is an important and practical application.
\end{abstract}

\section{Introduction} % [dpwe]}
\label{sec:intro}

% \begin{itemize}
%     \item Sound Scene Recomposition: Generative sound scene editing
%     \begin{itemize}
%         \item specific changes to existing complex scene
%         \item object-oriented editing
%     \end{itemize}
%     \item Recomposition vision: Event-roll-based editing
%     \begin{itemize}
%         \item Focus on deletion, enhancement, insertion
%         \item (transformation, translation, etc left for later)
%     \end{itemize}
%     \item UI: words + time specification
%     \begin{itemize}
%         \item Present user with event-level analysis of input (not the focus of this paper)
%         \item Command (class + action) + “activity-roll” conditioning
%     \end{itemize} 
% \end{itemize}

There are many scenarios in which an existing audio recording is edited to make small improvements.  These can range from global changes (e.g., frequency equalization, background noise removal) to local tweaks (removing a cough, making a doorbell more prominent).  Traditional audio editing software allows direct, explicit modification of particular parts of the waveform, but new {\em generative audio} techniques suggest a whole new level of capabilities -- for instance, allowing the ``filling-in'' of previously-obscured gaps in the soundtrack based on inference and large-scale models of general audio.

\begin{figure}[t]
  \centering
  %\centerline{\includegraphics[width=\columnwidth]{recompositron-ui.png}}
  \centerline{\includegraphics[width=\columnwidth]{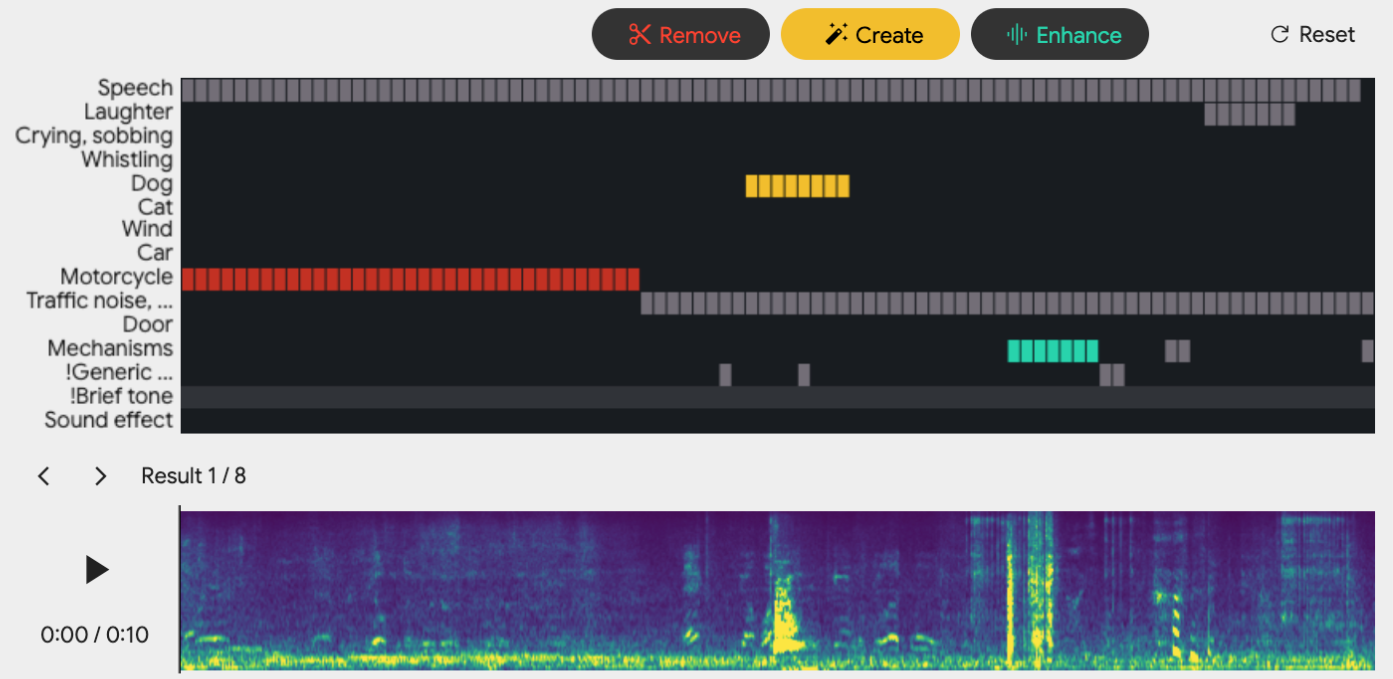}}
  \caption{Sound Recomposition editor shows the timing and inferred class of events in a sound scene.  The user can edit or insert individual events.}
  \label{fig:recompsition_ui}
\end{figure}

By considering a complex, real-world sound scene as a collection of individual sound events, many editing operations can be described as choosing particular sound events to modify (e.g., add, remove, alter) while holding the remainder of the scene unchanged.  These can be difficult with conventional sound editors owing to event overlap, but they become more feasible and natural in generative analysis-synthesis systems.  Figure~\ref{fig:recompsition_ui} illustrates a prototype interface: An {\em event-roll} showing individual sound events in the input is used as the control interface, allowing deletion, enhancement, and insertion of existing or novel sound events identified by class labels.

Here, we consider the problem of generating the edited audio output given the original audio and a set of text instructions paired with explicit time extents, which we term an {\em activity roll} (i.e., an event roll extended with actions).  For instance, the instruction could be "Delete dog bark" associated with time cells from 2.3 to 2.8 s.  
We do not, in this paper, address the sound scene event recognition required to build the event roll; this can be provided by existing Sound Event Recognition systems such as \cite{gong2024listenthinkunderstand}.  

We present a generative audio model able to modify individual sound events, trained on synthetically-constructed background-plus-foreground-event scenes.  These models are able to learn to remove, enhance, and insert sound events, and the activity roll representation provides precise and intuitive temporal control.  We report ablation experiments to illustrate the importance of the different edit controls -- action, target event class, and time extent.

The contributions of this work include:
\begin{itemize}
    \item The overall scenario of editing complex sound scenes at the level of perceived sound events, and the event roll user interface.
    \item Our realization of sound scene editing via an autoregressive generative audio transformer.
    \item Our ablation experiments showing the extent to which different components of the conditioning are vital to system performance.
\end{itemize}

% Specific discussion and contrasting of Delete, Insert, Enhance.
We focus on the three edit operations of Delete, Insert, and Enhance because of their practical value and the feasibility of constructing synthetic training examples (see section \ref{sec:training_data}).  These actions involve different model capabilities: Deletion involves the known ability of models to remove individual sound sources while reconstructing a coherent background (e.g., \cite{kilgour2022textdrivenseparationarbitrarysounds}).  Insertion of a given class category is a conventional Text-to-Audio task. Enhancement involves both source separation (to identify the weak audio event) and conditional generation (to add unobserved detail to the event regenerated at a higher amplitude).

% ronw: saving space
% Section \ref{sec:prev} reviews previous work in generative editing and sound scene analysis.  Section \ref{sec:approach} describes our approach, both to the model design and development, and to the synthesis of training data.  Section \ref{sec:evaluation} describes our evaluation process and results, including the ablation studies.  Section \ref{sec:discussion} discusses the implications and other ideas, and section \ref{sec:conclusion} gives our conclusions.

\section{Related Work} % [efonseca]}
\label{sec:prev}

% Previous work includes AUDIT~\cite{AUDIT_NEURIPS2023}

% \begin{itemize}
%     \item Generative audio
%     \item Editing content
%     \begin{itemize}
%         \item Editing audio
%     \end{itemize}
%     \item (Event-roll transcription)
% \end{itemize}

% TTA
Text-to-Audio models are capable of generating plausible general soundscapes from textual descriptions of the sound scenes \cite{kreuk2023audiogen,liu2023audioldm,ghosal2023text,liu2024audioldm}.
%Controllability
After the initial wave of Text-to-Audio models, and following the trend in computer vision \cite{zhang2023adding,zhao2023uni,sheynin2024emu}, a body of literature emerged putting emphasis on controlling the generation or modification of specific sound components in a sound scene or music sample \cite{AUDIT_NEURIPS2023,wu2024music,zhang2024instruct,garcia2024sketch2sound,wang2024audiocomposer}.

%Audio editing
In particular, unlike standard Text-to-Audio models, audio editing generative models often rely on edit instructions, usually in the form \textit{action} + \textit{class} (e.g., \textit{Enhance Laughter}) to enable concrete edits of the input audio.
One of the first audio editing works for general audio, AUDIT~\cite{AUDIT_NEURIPS2023}, leverages synthetic triplet data in the form (edit instruction, input audio, output audio) to train a diffusion model to perform edit tasks.
However, temporal control is poor as the time conditioning is solely based on natural language structures such as “at the beginning/middle/end”.
Our work improves this with the event-roll specification to allow millisecond-level temporal control of edits.

%PicoAudio
Our strategy for incorporating temporal control into an autoregressive model is conceptually similar to that recently proposed in \cite{xie2024picoaudio} for a diffusion model.
However, our \textit{activity roll} is not tied to a predefined fixed vocabulary, thereby supporting more flexibility than the timestamp matrix in \cite{xie2024picoaudio}.
Additionally, we leverage more comprehensive training/evaluation data (over 25$\times$ the target sound examples across more than twice as many sound classes).  We also report detailed ablations to demonstrate the importance of different conditioning components.
%\TODO{For reference, they use 636 examples across 18 classes.}

\section{Approach}
\label{sec:approach}

%\subsection{Model} % [ronw]}
%\label{sec:model}

\begin{figure}[t]
\vspace{-1mm}

  \centering
  % This just a first draft.
  % Source: http://docs/drawings/d/1mN4E9zmY4OFTOTMYOfBjeG2ZmoybSSvPZ_oc7uqmGXo?resourcekey=0-eZZulSM1hpEvhKXuuB2t2A
  \centerline{\includegraphics[width=1.0\columnwidth]{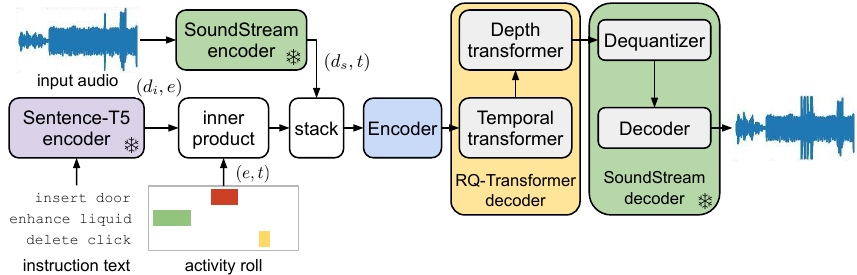}}
  \vspace{-1mm}
  \caption{Block diagram of the Recomposer model architecture.
  The network is an encoder-decoder transformer, taking a time-aligned stack of audio encodings and edit instruction embeddings as input, and autoregressively generating a SoundStream token sequence at the output.
  These tokens are passed through a SoundStream decoder to generate the output audio waveform.
  The text encoder (purple) and audio encoder/decoders (green) are pretrained and frozen.
  }
  \label{fig:block_diagram}
\end{figure}

Editing discrete events in audio scenes is simpler than full generative audio continuation~\cite{borsos2023audiolm} or text-to-audio generation~\cite{kreuk2023audiogen} given the strong conditioning of the input audio.
This allows the direct modeling of low level audio features without  hierarchical generation of intermediate ``semantic embeddings''~\cite{borsos2023audiolm,kreuk2023audiogen}.
The Recomposer model operates directly on a  SoundStream~\cite{zeghidour2021soundstream} representation of general audio, leveraging the codec's residual VQ (RVQ) structure to enable efficient autoregressive generation.  The overall Recomposer structure of an encoder-decoder~\cite{bahdanau2014neural} transformer~\cite{vaswani2017attention} is illustrated in Figure~\ref{fig:block_diagram}.  

The encoder input is created by stacking the continuous SoundStream $(d_s,t)$-shaped encoding of the input audio to a time-aligned representation of the set of edits to be applied.
The text describing individual edit instructions is encoded with a pretrained Sentence-T5~\cite{ni-etal-2022-sentence} network, which
encodes a variable-length text token sequence into a single $d_i$-dimensional vector, resulting in one instruction embedding for each of $e$ edits in a $(d_i,e)$-shaped matrix.
These embeddings are extended over the $t$-step time axis by taking the inner product with the corresponding $(e,t)$-shaped binary activity roll, time-aligned with the audio encoding, to create an overall $(d_i,t)$ instruction matrix.
%The resulting time-aligned editing instruction matrix is stacked with a continuous-valued encoding of the input sound computed by passing the input waveform through a pretrained SoundStream encoder.
 
Feeding the stacked input matrix into an encoder-decoder RQ-Transformer~\cite{lee2022autoregressive} generates a sequence of quantized SoundStream RVQ tokens representing the output.
The %RQ-Transformer
decoder, adapted to audio similarly to \cite{defossez2024moshi}, is divided into two autoregressive subnetworks: a temporal transformer which generates an encoding for each output frame, conditioned on the previous frame; and a depth transformer, which generates each RVQ token in sequence, conditioned on the temporal transformer encoding and the preceding RVQ tokens in the same frame.
The resulting RVQ tokens are converted back into waveforms using a pretrained SoundStream dequantizer and decoder.
%
%  \item Audio used to train the SoundStream model?  % looks like  soundstream_16khz_generalaudio, which was trained on "Librivox, Librivox + Freesound noise, Magnatagatune"
We use the same SoundStream model at input and output, trained on a variety of audio content, including LibriVox speech samples mixed with non-speech background sampled from Freesound~\cite{font2013freesound,Mixit_NEURIPS2020}, and music from %MagnaTagATune~
\cite{law2010evaluation}, at 16~kHz sample rate.

The transformer architecture follows the convention of BERT~large~\cite{devlin2019bert}, using 12 layer encoder, 12 layer temporal transformer, and an additional 3 layers in the depth transformer, resulting in a total of about 390M trainable parameters.
        
%    \begin{itemize}
%        \item Text input: Sentence-T5~\cite{ni-etal-2022-sentence} embedding
%        \item SoundStream~\cite{zeghidour2021soundstream} input, tokenized soundstream output 
%        \item Encoder-decoder \cite{bahdanau2014neural,vaswani2017attention} architecture
%    \end{itemize}

% Should be move this into the following section and rename it "Experiments"?
\subsection{Training data}% [dpwe]}
\label{sec:training_data}

\begin{figure}[t]
\vspace{-1mm}

  \centering
  \centerline{\includegraphics[width=\columnwidth]{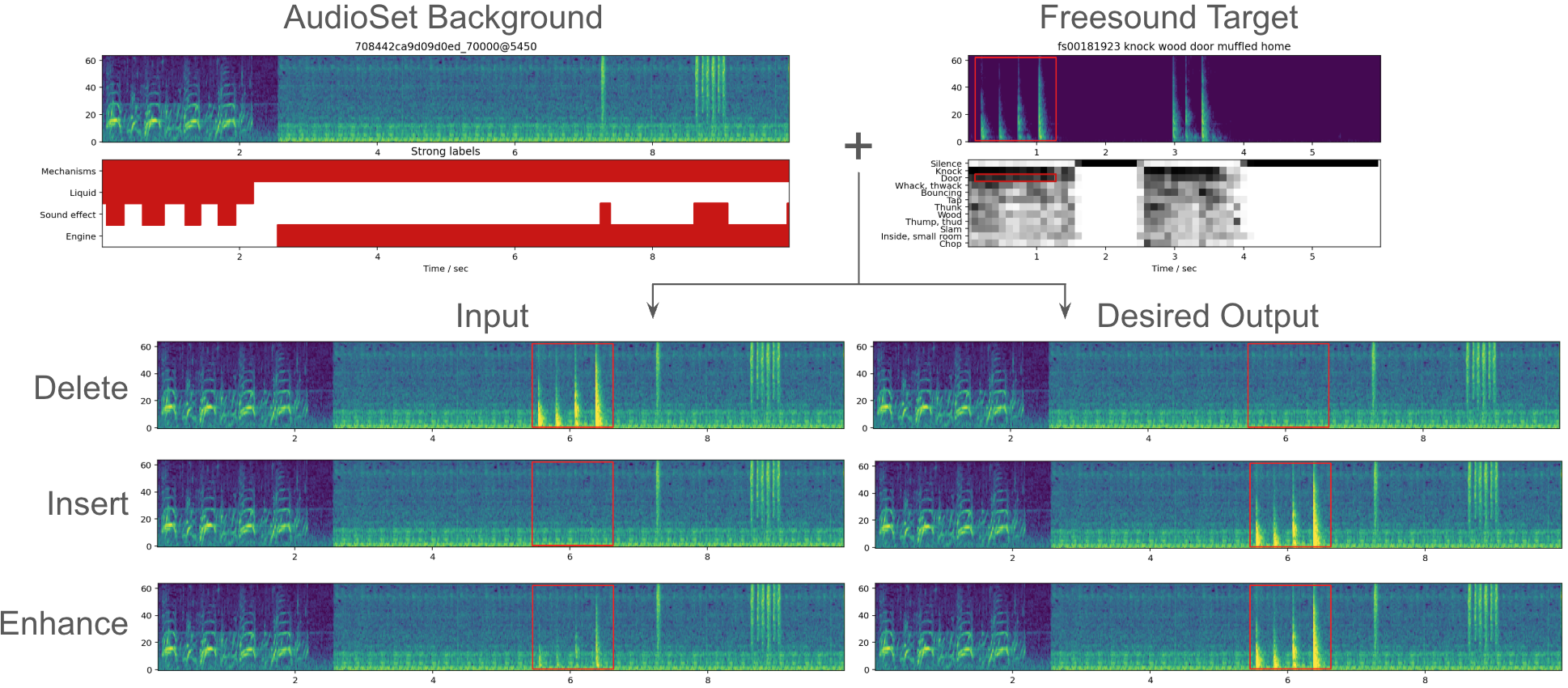}}
  \vspace{-1mm}
  \caption{Synthetic data creation: Backgrounds are chosen from AudioSet based on the of variety and density of their human annotations.  Targets come from Freesound, via keyword match and classifier and energy envelope refinement (the red box).  Mixing target and background at random time offset and various local target-to-background ratios gives input and desired output training pairs for edit operations Delete, Insert, and Enhance.}
\label{fig:synthetic_data}
\end{figure}

A sufficient volume of training examples is crucial to any deep learning system.  To generate pairs of input sound scenes and desired modified output scenes, we employ synthetic mixing of individual `target' sound events with real, dense `background' sound scenes.  Synthetic mixing allows us to provide exact examples of backgrounds with and without the target (for Delete, and, when flipped, for Insert), and mixing the target at different levels provide examples for Enhancement.

The process of data generation is illustrated in Figure \ref{fig:synthetic_data}.  We drew backgrounds from the strong-labeled portion of AudioSet \cite{hershey2021benefittemporallystronglabelsaudio}.  These diverse sound scene excerpts are filtered based on the temporally-precise labels to avoid examples with large silent gaps (i.e., no more than 10\% of the timeline unlabeled) and to ensure diversity of content (i.e., at least two different sound classes present). This yielded 167,961 background clips of 10 s each for training.

Target events are drawn from Freesound \cite{font2013freesound}, which has a much higher incidence of single-source sound examples, suitable as isolated foregrounds.  We collect examples of events in specific classes by (a) retrieving Freesound clips based on matching keyword tags (``door'' in Figure \ref{fig:synthetic_data}), (b) using an audio event classifier \cite{yamnet} to identify specific time ranges that reflect the intended class.  We use hysteresis thresholding of the corresponding classifier output scores (at a 10 Hz sample rate) to find compact example events, then further refine their time extent using the sound's energy envelope relative to an estimated noise floor; only examples well above the floor are used.  Target events must have duration between 0.2 and 2.0 s.  This resulted in 16,366 training examples spread across 40 AudioSet classes, chosen to be short, well-defined sound events.

Synthetic mixtures are constructed by mixing random events with random backgrounds at random times.  The mix level of the target is controlled via the target-to-background energy ratio or TBR:  We calculate the energy of the target event waveform, the energy of just the overlapping region of the background, then scale the target to achieve the intended TBR.  We used TBR of 10~dB for input events to be Deleted; -6~dB for input events to be Enhanced; and 10~dB for desired output events to make them visible in metrics.  To promote generalization we roved the input target levels $\pm 3$~dB.  % and $\pm0.5$~dB for outputs.

Our training data was synthesized on-the-fly without repeating.  Our primary data, referred to as ``EDIN'' (for enhancement/deletion/insertion/no-op), gave each example two `edits'  independently and uniformly chosen from the set {enhance, delete, insert, no-op}; thus, 6.25\% ($0.25 * 0.25$) of examples received two no-ops, resulting in a desired output the same as the input. For evaluation, we generated frozen synthetic mixtures, including separate test sets of 10,000 examples for each edit operation.  The backgrounds and targets for the evaluation sets were drawn from distinct pools based on the AudioSet and FSD50K \cite{fonseca2022FSD50K} eval sets, providing 24,098 background clips and 1,697 target events.

For a separate Enhancement-only model, we trained with examples all containing exactly one Enhance edit, but with a broader TBR range chosen uniformly between -30 dB and 0 dB.  Models trained on this Enhancement data are evaluated on separate Enhance eval sets for each input event TBR level from -30 dB to 0 dB in steps of 3 dB.

From informal listening, the mixtures sounded fairly natural.  However, the unrelated recording conditions of target and background could give unnatural cues to help the model identify the target portion.  Additionally, there was no effort to make the target event semantically match the background, so some mixtures are clearly incongruous.

\section{Evaluation}% [dpwe]}
\label{sec:evaluation}

%\subsection{Metrics}
%\label{sec:metrics}

We use two metrics to evaluate our model-generated estimate waveforms against the synthetic desired outputs:
\begin{itemize}
    \item {\bf Multiscale Spectral Distortion (MSD)} \cite{wang2019neural,engelddsp} calculates the cellwise difference between spectrograms calculated at various time resolutions, averaging both linear- and log-domain results.  MSD is essentially a signal-level metric, but is more tolerant of minor differences in timing when compared to waveform mean-squared error. It is thus most informative for conditions where the model has a chance of precisely predicting the desired output waveform, as in Deletion, or Enhancement when the input target is excessively weak.
    \item {\bf Classifier KL Divergence (KLD)} following previous work \cite{yang2023diffsound,kreuk2023audiogen,liu2023audioldm,AUDIT_NEURIPS2023} applies a sound event classifier to both waveforms and calculates the Kullback-Liebler divergence between distributions across classes, normalized to be posteriors.  We used the open-source YAMNet \cite{yamnet}.  By viewing the signal through the lens of a classifier, we can ignore waveform details that do not change the inferred class.  This makes KLD a useful metric for operations such as Insertion and Enhancement of very quiet events, both of which involve significant ``generation'' of new signal conditioned on the event class description.  
\end{itemize}
Both metrics are calculated per time frame, making them sensitive to the temporal structure of the sound scenes -- important in this work that deals specifically with event timing.  This precision allows us to calculate the average per-frame values over both the target time region (i.e., where the target event has been added or removed), and the remainder (which is ideally unchanged by processing).
We don’t use Fr{\'e}chet audio distance \cite{kilgour2019frechetaudiodistancemetric} because it doesn't support such fine-grained measurements, and because recent work has questioned its correlation with perceptual quality  \cite{gui2024adapting,tailleur2024correlation}.
%We don't use FAD \cite{kilgour2019frechetaudiodistancemetric} for various reasons, including its inability to provide this level of precision.

Metrics are calculated after passing all waveforms through the SoundStream codec which had minimal perceptual impact.  Without this, codec distortion would have largely swamped the effects of the model for both metrics.  %Calculating metrics in the post-codec domain exposes the relevant model behavior.

%[dpwe] Figure illustrating framewise metric calculation?

\subsection{Results}
\label{sec:results}

%% Table captions should be placed above the table, as in \cref{tab:table1}. We highly encourage the usage of the \texttt{booktabs} package for pretty tables and the \texttt{siunitx} package to properly format and align numbers in table columns.

\begin{table}[t]
\centering
\caption{Evaluation results for the unablated model using separate single-target-event evaluation sets for
each editing action, Delete, Insert, and Enhance.  The metrics calculated from comparing the model inputs to the desired outputs are reported as a 
no-processing baseline.  For each condition, we report average per-frame metrics separately for the target and nontarget time ranges. Lower is better for both metrics, with 0 indicating perfect match.}
\label{tab:results}
%\sisetup{
%    %reset-text-series = false, 
%    %text-series-to-math = true, 
%    mode=text,
%    tight-spacing=true,
%    round-mode=places,
%    round-precision=1,
%    table-format=2.2,
%    table-number-alignment=center
%}
\setlength{\tabcolsep}{1.5ex}
%\begin{tabular}{@{}ll*{3}{S[round-precision=1,table-format=2.1]S}@{}}
%\begin{tabular}{@{}ll*{3}{S[round-precision=1,table-format=2.1]@{\hskip1ex}S}@{}}
\begin{tabular}{@{}ll*{3}{c@{\hskip1ex}c}@{}}
    \toprule
    & & \multicolumn{2}{c}{Delete} & \multicolumn{2}{c}{Insert} & \multicolumn{2}{c}{Enhance}\\
    \cmidrule(lr){3-4}\cmidrule(lr){5-6}\cmidrule(l){7-8}
    Region    & Signal   & {MSD}& {KLD} & {MSD}& {KLD} & {MSD}& {KLD}\\
    \midrule
    Target    & input    &  4.8 & 1.6 &  4.8 & 2.8 &  3.4 & 1.6  \\
              & estimate &  2.5 & 0.5 &  5.1 & 1.9 &  2.6 & 0.9  \\
    \addlinespace
    Nontarget & input    &  0.0 & 0.0 &  0.0 & 0.0 &  0.0 & 0.0  \\
              & estimate &  1.3 & 0.3 &  1.3 & 0.3 &  1.3 & 0.3  \\
    \bottomrule
\end{tabular}
\end{table}

Table \ref{tab:results} gives the main results for the general-purpose model trained on the EDIN data, with up to two edits per sample. The table shows results for per-action eval sets, reporting the metrics separately for the target and nontarget time ranges, and contrasting the model output estimates with the model inputs (where both are compared to the desired outputs, and all comparisons are made on waveforms reconstructed from the SoundStream encoding).  In the target regions, the model estimates show consistent improvements over the unprocessed condition, with the largest MSD improvement of 2.3 for Delete, and the smallest for Insert (0.3); since Insert MSD is comparing a true example of the requested class with one generated by the model based only on class name and timing, any improvement is welcome.  For KLD, Insert gives the greatest divergences, but also large improvements almost as good as from Delete, validating that the model's generated events are recognized as resembling the class present in the desired output.  In the nontarget regions, we see that the input exactly matches the desired output for all conditions (metrics of 0.0).  The model output estimates show a fixed, but constant, offset reflecting the distortion introduced by copying the input through the model.

\begin{figure}[t]
% \vspace{-1mm}
  \centering
  \centerline{\includegraphics[width=\columnwidth]{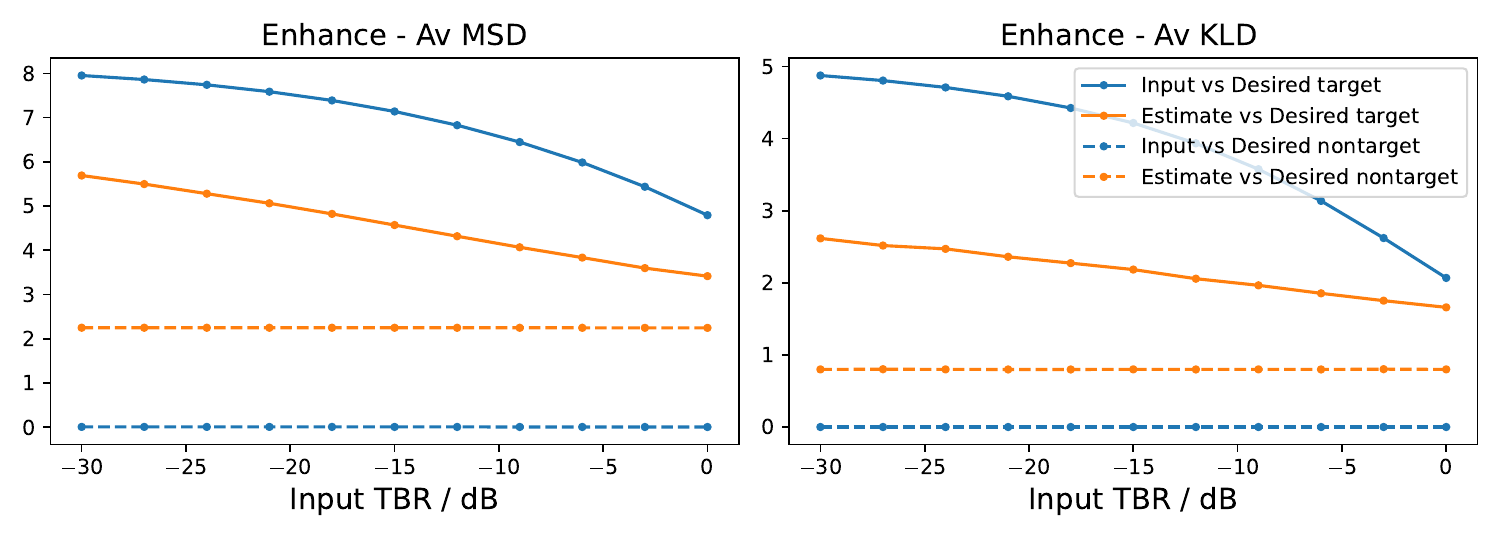}}
  \vspace{-2mm}
  \caption{MSD and KLD eval results for Enhancement as a function of input TBR.  Orange lines compare model estimates to desired targets; blue lines compare model inputs to desired targets.  Solid lines are average over target (modified) regions; dashed lines average over nontarget (unmodified) regions.}
\label{fig:enhancement_results}
\end{figure}

Figure \ref{fig:enhancement_results} shows the results of an experiment to investigate how Enhancement transitions from lightly-guided audio event generation for very weak inputs, to something closer to source separation when the target input is more clearly discernable. A single model was trained for enhancement only with a mix of input TBRs ranging over -30 to 0~dB; the desired output signals always had the target at 15~dB TBR to minimize the model's uncertainty about what was expected.  We see that both the MSD and KLD for the target-region model outputs estimate vs. desired improve steadily with increasing input TBR, showing consistent substantial improvements over no-processing (input vs. desired); nontarget regions again show a small, constant distortion due to model processing. The best MSD improvement relative to no-processing occurs for intermediate TBRs of around -15 dB.  It's interesting that KLD improves with TBR, since the classifier might uniformly reflect that the model had  generated a target of the same class as in the desired output across all input target levels.  However, the YAMNet classifier only reduces, but does not eliminate, the tendency of the metric to prefer a close match in waveform, not just in class identity. 

\begin{figure}[t]
  \centering
  \centerline{\includegraphics[width=\columnwidth]{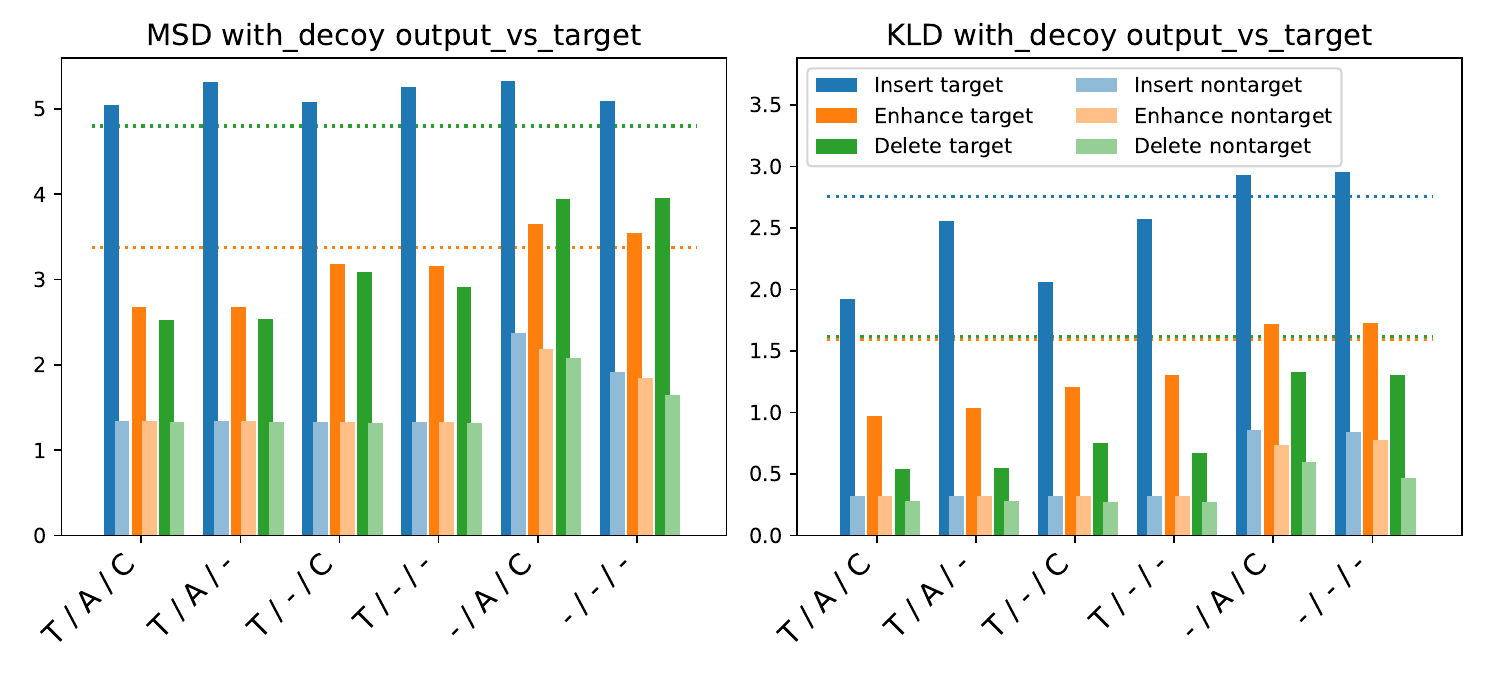}}
  \vspace{-3mm}
  \caption{Evaluation results for the ablation study on the decoy events dataset. X axis label indicates the conditioning information provided; T:~timing, A:~action, C:~class.
  Bars compare models estimates to desired outputs for each edit operation, with darker bars reflecting target regions and lighter bars for non-target regions. Dotted lines show the baseline of comparing the unmodified input signal to the target, averaged over the target regions; The Enhance baseline is lower for MSD because of the attenuated target in the input, and the Insert baseline is higher for KLD because of the asymmetry in KL divergence.  (Distance between inputs and targets is zero outside of target regions, so is not plotted.)}
  \label{fig:ablation_results}
\end{figure}

To understand the value of the different edit conditioning information -- namely, the requested action, the class identity specified for the target, and the precise timing from the activity roll -- we conducted ablation experiments whose results are shown in Figure \ref{fig:ablation_results}.  We trained 6 different models in which different parts of the conditioning were ablated.  To make it harder for the model to guess the intended operation, Insert and Enhance evaluation examples include a second copy of the target event at a non-overlapping time.  Its level is the same in both input and desired output, so ideally it will not affect the metrics, but if the model is ignoring the timing conditioning (or that information is ablated, as in the final two conditions), the model may erroneously process the decoy.

The results are broadly in line with expectations:  All ablations lead to worse metrics for the target region, whereas the nontarget regions are only affected when the timing information is ablated (forcing the model to guess which event to process, or where to insert).  Insertion is particularly hard-hit when the target class identity is omitted, moreso when measured by the classifier-based KLD, whereas Deletion and Enhancement show little or no benefit from being told the target class, but gives worse results when the edit action is ablated (that is, the evaluation set consisting of examples that are intended to be deleted/enhanced appear to result in fewer of the intended edits when the model is forced to guess).  Enhancement examples show cumulative benefit from both action and class information, most visibly in the KLD.

\subsection{Subjective impressions}
\label{sec:impressions}
    
    %Audio examples from base model and ablations here: http://colab/drive/1zsyUx-Zu_p_uuvn1ZaSwWUUZvTJrOKcF?resourcekey=0-IDOD1B_lPIw8hWDFdMuCLQ#scrollTo=1gFqWHffMyS2
    %${\it$<$See .zip file in supplementary materials; Example audio will be made publicly available upon publication.$>$}. 
    Informal listening\footnote{Audio examples can be heard at: \url{https://storage.googleapis.com/recomposer/index.html}} yielded impressions consistent with our quantitative findings.  Audio from unmodified regions is passed through with little distortion.  Deleted events are usually fully deleted, and replaced by reasonable background audio.  Insertion and Enhancement generally work well.  As expected, the inserted/enhanced events are of the correct sound class, though for enhancement they sometimes sound somewhat different from the pre-enhancement input event.  When insertions and deletions fail, they tend to do nothing rather than a partial edit.
    
    With timing information ablated, Delete decoy eval sets may have both events or neither deleted, and sometimes sounds from unrelated classes are inserted.  For Insert and Enhance, usually some time-localized part of the clip is modified, but usually not at the desired time and often not with the correct sound class.  
    
    When sound class is ablated (but timing information and edit type are given), deletions and enhancements usually succeed because the model can identify the target event from timing alone.  For insertions, a sound from an unrelated class is usually inserted at the correct time.
    
    When edit type is ablated (but timing information and sound class are given), both enhancements and deletion examples are typically subjected to enhancement edits.  Insertion works well since, in our data, the fact that the requested edit is insertion can be inferred from the fact that there's no prominent sound event in the input at the specified time.

\section{Discussion}
\label{sec:discussion}

Although the Recomposer model addresses several realistic editing scenarios, we recognize that it is more of a proof-of-concept than a practical tool.  Using the interface illustrated in Figure \ref{fig:recompsition_ui} we have interactively investigated the model's performance on non-synthetic sound scenes (but with ground-truth event transcripts, needed for the interface).  By training on examples with between zero and two edits, the model is able to perform several edits in a single application.  The relatively wide range of input event level roving in the training data allows the deletion and enhancement of many sound events.  However, the generated output events are always at the high 15dB TBR used in training; a practical system would need more mechanisms for specifying generated event properties.

The current vocabulary of event descriptions is strictly limited to the subset of fixed AudioSet class labels.  To support richer changes, perhaps including transformation of events, we would like to be able to train with much more diverse and detailed text descriptions of events.  Constructing these training materials is a separate and challenging problem, however.  The existing Sentence-T5 encoder should be directly usable with richer descriptions.

Additional conditioning could improve utility.  Very often, soundtracks to be edited are associated with video, and generating soundtracks that are inconsistent with the visuals is not useful.  Adding video-derived conditioning for the generation could avoid this, for a video-plus-edits-to-audio generation system.  % Reference V2A work?
Edit specifications could also include other, non-time-local guidance, such as changes to the overall acoustic environment.

Our objective evaluations confirmed the expected impacts of different conditions, but the measures were also sensitive to distortion intrinsic to model processing that did not seem perceptually significant.  Subjectively, the important features of individual model estimates, aside from their general success in reproducing the input audio, were whether the correct edit was applied, and whether it affected the correct perceived source in the original mixture.  It would be possible to design metrics that attempted to measure this more directly, for instance by characterizing the difference between input and model output as primarily concentrated in a particular time range (rather than using the oracle target time range as in our metrics).

% \begin{itemize}
%     \item It works
%     \item Ablations highlight risk of model outwitting experimenter
%     \item Insertion-Enhancement-Transformation
%     \item Multiple edits
%     \item Other kinds of edits
%     \begin{itemize}
%         \item Within-class attributes - “make it more aggressive” - training data
%     \end{itemize}
%     \item Where to get the ground-truth event roll?
%     \begin{itemize}
%         \item Improved time-sensitive classifiers
%     \end{itemize}
%     \item Where to get training data?
%     \begin{itemize}
%         \item Separate-then-remix for more natural examples
%         \begin{itemize}
%             \item Model can learn dependence of insertions on context
%         \end{itemize}
%     \end{itemize}
%     \item Text conditioning
%     \begin{itemize}
%         \item Normalize through LLM
%     \end{itemize}
%     \item Other conditioning
%     \begin{itemize}
%         \item Visual input
%         \item General text
%         \item Broader representation of target audio domain
%     \end{itemize}
%     \item Evaluation challenges
%     \begin{itemize}
%         \item Did it get “class” right?
%         \item Naturalness of insertions?
%         \item Limitations even of human evaluation?
%     \end{itemize}
%     \item enhancement as the main story - cross-enhancement
% \end{itemize}

\section{Conclusion}
\label{sec:conclusion}

We introduce Recomposer, an approach to sound-event-oriented editing of real-world sound scenes.  We show that synthetic, but realistic, pairs of input and desired output waveforms -- illustrating the edit actions Delete, Insert, and Enhance -- can be used to train an encoder-decoder transformer that subjectively succeeds at making the intended edits, at least in many examples.  Our model uses a combination of explicit timing represented as a vector of time-frame flags, and vector-encoding of edit instructions (actions plus class descriptions) derived, in principle, from free text.  While the model lacks some features needed to make it truly useful (including a limited vocabulary of event descriptions), we feel it shows the feasibility of the approach.  Future work most likely hinges on richer and more sophisticated approaches for generating training data of ({\em input waveform, desired output waveform, edit description and timing}) tuples.

% \begin{itemize}
%     \item Generative editing paradigm
%     \begin{itemize}
%         \item Pure-text interface not the ideal match
%     \end{itemize}
%     \item Works fine, training data is main determinant
%     \item Useful part of a future AI-enhanced content editing workflow
% \end{itemize}

\section*{Acknowledgments}
% Acknowledge: mtagliasacchi@, zborsos@, bmcwilliams@ for depthformer, getreuer@ for UI, etzinis@ for data

Thanks to Marco Tagliasacchi, Zal\'{a}n Borsos, and Brian McWilliams for the depth transformer SoundStream generation.  Thanks to Matt Harvey for pilot work on synthetic targets and the Freesound infrastructure, Pascal Getreuer for the Recomposer interface of Figure \ref{fig:recompsition_ui}, and Efthymios Tzinis for providing additional training materials.  Many thanks to Frederic Font of UPF for helping with access to Freesound.

\bibliographystyle{IEEEtran}
\bibliography{refs25}

% Generated by IEEEtran.bst, version: 1.14 (2015/08/26)
\begin{thebibliography}{10}
\providecommand{\url}[1]{#1}
\csname url@samestyle\endcsname
\providecommand{\newblock}{\relax}
\providecommand{\bibinfo}[2]{#2}
\providecommand{\BIBentrySTDinterwordspacing}{\spaceskip=0pt\relax}
\providecommand{\BIBentryALTinterwordstretchfactor}{4}
\providecommand{\BIBentryALTinterwordspacing}{\spaceskip=\fontdimen2\font plus
\BIBentryALTinterwordstretchfactor\fontdimen3\font minus
  \fontdimen4\font\relax}
\providecommand{\BIBforeignlanguage}[2]{{%
\expandafter\ifx\csname l@#1\endcsname\relax
\typeout{** WARNING: IEEEtran.bst: No hyphenation pattern has been}%
\typeout{** loaded for the language `#1'. Using the pattern for}%
\typeout{** the default language instead.}%
\else
\language=\csname l@#1\endcsname
\fi
#2}}
\providecommand{\BIBdecl}{\relax}
\BIBdecl

\bibitem{gong2024listenthinkunderstand}
Y.~Gong, H.~Luo, A.~H. Liu, L.~Karlinsky, and J.~Glass, ``Listen, think, and
  understand,'' in \emph{Proc. ICLR}, 2024.

\bibitem{kilgour2022textdrivenseparationarbitrarysounds}
K.~Kilgour, B.~Gfeller, Q.~Huang, A.~Jansen, S.~Wisdom, and M.~Tagliasacchi,
  ``Text-driven separation of arbitrary sounds,'' in \emph{Proc. Interspeech},
  2022.

\bibitem{kreuk2023audiogen}
F.~Kreuk, G.~Synnaeve, A.~Polyak, U.~Singer, A.~D{\'e}fossez, J.~Copet,
  D.~Parikh, Y.~Taigman, and Y.~Adi, ``{AudioGen}: Textually guided audio
  generation,'' in \emph{Proc. ICLR}, 2023.

\bibitem{liu2023audioldm}
H.~Liu, Z.~Chen, Y.~Yuan, X.~Mei, X.~Liu, D.~Mandic, W.~Wang, and M.~D.
  Plumbley, ``{AudioLDM}: Text-to-audio generation with latent diffusion
  models,'' in \emph{Proc. ICML}, 2023, pp. 21\,450--21\,474.

\bibitem{ghosal2023text}
D.~Ghosal, N.~Majumder, A.~Mehrish, and S.~Poria, ``Text-to-audio generation
  using instruction guided latent diffusion model,'' in \emph{Proc. ACM
  Multimedia}, 2023, pp. 3590--3598.

\bibitem{liu2024audioldm}
H.~Liu, Y.~Yuan, X.~Liu, X.~Mei, Q.~Kong, Q.~Tian, Y.~Wang, W.~Wang, Y.~Wang,
  and M.~D. Plumbley, ``{AudioLDM} 2: Learning holistic audio generation with
  self-supervised pretraining,'' \emph{IEEE/ACM Trans. Audio, Speech, Lang.
  Process.}, 2024.

\bibitem{zhang2023adding}
L.~Zhang, A.~Rao, and M.~Agrawala, ``Adding conditional control to
  text-to-image diffusion models,'' in \emph{Proc. CVPR}, 2023, pp. 3836--3847.

\bibitem{zhao2023uni}
S.~Zhao, D.~Chen, Y.-C. Chen, J.~Bao, S.~Hao, L.~Yuan, and K.-Y.~K. Wong,
  ``{Uni-ControlNet}: All-in-one control to text-to-image diffusion models,''
  \emph{Advances in Neural Information Processing Systems}, vol.~36, pp.
  11\,127--11\,150, 2023.

\bibitem{sheynin2024emu}
S.~Sheynin, A.~Polyak, U.~Singer, Y.~Kirstain, A.~Zohar, O.~Ashual, D.~Parikh,
  and Y.~Taigman, ``{Emu Edit}: Precise image editing via recognition and
  generation tasks,'' in \emph{Proc. CVPR}, 2024, pp. 8871--8879.

\bibitem{AUDIT_NEURIPS2023}
Y.~Wang, Z.~Ju, X.~Tan, L.~He, Z.~Wu, J.~Bian, and S.~Zhao, ``{AUDIT}: Audio
  editing by following instructions with latent diffusion models,'' in
  \emph{Advances in Neural Information Processing Systems}, vol.~36, 2023, pp.
  71\,340--71\,357.

\bibitem{wu2024music}
S.-L. Wu, C.~Donahue, S.~Watanabe, and N.~J. Bryan, ``{Music ControlNet}:
  Multiple time-varying controls for music generation,'' \emph{IEEE/ACM Trans.
  Audio, Speech, Lang. Process.}, vol.~32, pp. 2692--2703, 2024.

\bibitem{zhang2024instruct}
Y.~Zhang, Y.~Ikemiya, W.~Choi, N.~Murata, M.~A. Mart{\'\i}nez-Ram{\'\i}rez,
  L.~Lin, G.~Xia, W.-H. Liao, Y.~Mitsufuji, and S.~Dixon,
  ``{Instruct-MusicGen}: Unlocking text-to-music editing for music language
  models via instruction tuning,'' \emph{arXiv preprint arXiv:2405.18386},
  2024.

\bibitem{garcia2024sketch2sound}
H.~F. Garc{\'\i}a, O.~Nieto, J.~Salamon, B.~Pardo, and P.~Seetharaman,
  ``{Sketch2Sound}: Controllable audio generation via time-varying signals and
  sonic imitations,'' \emph{arXiv preprint arXiv:2412.08550}, 2024.

\bibitem{wang2024audiocomposer}
Y.~Wang, H.~Chen, D.~Yang, Z.~Wu, H.~Meng, and X.~Wu, ``{AudioComposer}:
  Towards fine-grained audio generation with natural language descriptions,''
  in \emph{Proc. ICASSP}, 2025.

\bibitem{xie2024picoaudio}
Z.~Xie, X.~Xu, Z.~Wu, and M.~Wu, ``{PicoAudio}: Enabling precise timestamp and
  frequency controllability of audio events in text-to-audio generation,''
  \emph{arXiv preprint arXiv:2407.02869}, 2024.

\bibitem{borsos2023audiolm}
Z.~Borsos, R.~Marinier, D.~Vincent, E.~Kharitonov, O.~Pietquin, M.~Sharifi,
  D.~Roblek, O.~Teboul, D.~Grangier, M.~Tagliasacchi \emph{et~al.},
  ``{AudioLM}: a language modeling approach to audio generation,''
  \emph{IEEE/ACM Trans. Audio, Speech, Lang. Process.}, vol.~31, pp.
  2523--2533, 2023.

\bibitem{zeghidour2021soundstream}
N.~Zeghidour, A.~Luebs, A.~Omran, J.~Skoglund, and M.~Tagliasacchi,
  ``{SoundStream}: An end-to-end neural audio codec,'' \emph{IEEE/ACM Trans.
  Audio, Speech, Lang. Process.}, vol.~30, pp. 495--507, 2021.

\bibitem{bahdanau2014neural}
D.~Bahdanau, K.~Cho, and Y.~Bengio, ``Neural machine translation by jointly
  learning to align and translate,'' in \emph{Proc. ICLR}, 2015.

\bibitem{vaswani2017attention}
A.~Vaswani, N.~Shazeer, N.~Parmar, J.~Uszkoreit, L.~Jones, A.~N. Gomez,
  {\L}.~Kaiser, and I.~Polosukhin, ``Attention is all you need,''
  \emph{Advances in Neural Information Processing Systems}, vol.~30, 2017.

\bibitem{ni-etal-2022-sentence}
J.~Ni, G.~Hernandez~Abrego, N.~Constant, J.~Ma, K.~Hall, D.~Cer, and Y.~Yang,
  ``{Sentence-T5}: Scalable sentence encoders from pre-trained text-to-text
  models,'' in \emph{Proc. ACL}, May 2022.

\bibitem{lee2022autoregressive}
D.~Lee, C.~Kim, S.~Kim, M.~Cho, and W.-S. Han, ``Autoregressive image
  generation using residual quantization,'' in \emph{Proc. CVPR}, 2022.

\bibitem{defossez2024moshi}
A.~D{\'e}fossez, L.~Mazar{\'e}, M.~Orsini, A.~Royer, P.~P{\'e}rez,
  H.~J{\'e}gou, E.~Grave, and N.~Zeghidour, ``Moshi: a speech-text foundation
  model for real-time dialogue,'' \emph{arXiv preprint arXiv:2410.00037}, 2024.

\bibitem{font2013freesound}
F.~Font, G.~Roma, and X.~Serra, ``Freesound technical demo,'' in \emph{Proc.
  ACM Multimedia}, 2013, pp. 411--412.

\bibitem{Mixit_NEURIPS2020}
S.~Wisdom, E.~Tzinis, H.~Erdogan, R.~J. Weiss, K.~Wilson, and J.~Hershey,
  ``Unsupervised sound separation using mixture invariant training,'' in
  \emph{Advances in Neural Information Processing Systems}, vol.~33, 2020, pp.
  3846--3857.

\bibitem{law2010evaluation}
E.~Law, K.~West, M.~Mandel, M.~Bay, and J.~Downie, ``Evaluation of algorithms
  using games: the case of music annotation,'' in \emph{Proc. ISMIR}, 2010.

\bibitem{devlin2019bert}
J.~Devlin, M.-W. Chang, K.~Lee, and K.~Toutanova, ``{BERT}: Pre-training of
  deep bidirectional transformers for language understanding,'' in \emph{Proc.
  NAACL}, 2019.

\bibitem{hershey2021benefittemporallystronglabelsaudio}
S.~Hershey, D.~P.~W. Ellis, E.~Fonseca, A.~Jansen, C.~Liu, R.~Channing~Moore,
  and M.~Plakal, ``The benefit of temporally-strong labels in audio event
  classification,'' in \emph{Proc. ICASSP}, 2021, pp. 366--370.

\bibitem{yamnet}
\BIBentryALTinterwordspacing
M.~Plakal and D.~P.~W. Ellis, ``Sound classification with {YAMNet},'' 2020.
  [Online]. Available: \url{https://www.tensorflow.org/hub/tutorials/yamnet}
\BIBentrySTDinterwordspacing

\bibitem{fonseca2022FSD50K}
E.~Fonseca, X.~Favory, J.~Pons, F.~Font, and X.~Serra, ``{FSD50K}: an open
  dataset of human-labeled sound events,'' \emph{IEEE/ACM Trans. Audio, Speech,
  Lang. Process.}, vol.~30, pp. 829--852, 2022.

\bibitem{wang2019neural}
X.~Wang, S.~Takaki, and J.~Yamagishi, ``Neural source-filter waveform models
  for statistical parametric speech synthesis,'' \emph{IEEE/ACM Trans. Audio,
  Speech, Lang. Process.}, vol.~28, pp. 402--415, 2019.

\bibitem{engelddsp}
J.~Engel, L.~Hantrakul, C.~Gu, and A.~Roberts, ``{DDSP}: Differentiable digital
  signal processing,'' in \emph{Proc. ICLR}, 2020.

\bibitem{yang2023diffsound}
D.~Yang, J.~Yu, H.~Wang, W.~Wang, C.~Weng, Y.~Zou, and D.~Yu, ``Diffsound:
  Discrete diffusion model for text-to-sound generation,'' \emph{IEEE/ACM
  Trans. Audio, Speech, Lang. Process.}, vol.~31, pp. 1720--1733, 2023.

\bibitem{kilgour2019frechetaudiodistancemetric}
K.~Kilgour, M.~Zuluaga, D.~Roblek, and M.~Sharifi, ``Fr\'echet audio distance:
  A metric for evaluating music enhancement algorithms,'' in \emph{Proc.
  Interspeech}, 2019.

\bibitem{gui2024adapting}
A.~Gui, H.~Gamper, S.~Braun, and D.~Emmanouilidou, ``Adapting {Fr{\'e}chet}
  audio distance for generative music evaluation,'' in \emph{Proc. ICASSP},
  2024, pp. 1331--1335.

\bibitem{tailleur2024correlation}
M.~Tailleur, J.~Lee, M.~Lagrange, K.~Choi, L.~M. Heller, K.~Imoto, and
  Y.~Okamoto, ``Correlation of {Fr{\'e}chet} audio distance with human
  perception of environmental audio is embedding dependent,'' in \emph{Proc.
  EUSIPCO}, 2024, pp. 56--60.

\end{thebibliography}

\end{document}